\documentclass[pmlr]{jmlr}

\usepackage{longtable}

\usepackage{booktabs}

\usepackage[load-configurations=version-1]{siunitx}

\usepackage{caption}
\usepackage{svg}

\theorembodyfont{\upshape}
\theoremheaderfont{\scshape}
\theorempostheader{:}
\theoremsep{\newline}

\jmlrvolume{273}
\jmlryear{2025}
\jmlrworkshop{iRAISE at AAAI 2025}

\title[ARCHED: Human-Centered AI Framework for Instructional Design]{ARCHED: A Human-Centered Framework for Transparent, Responsible, and Collaborative AI-Assisted Instructional Design}

\author{
\Name{Hongming Li}$^{1}$ \Email{hli3@ufl.edu} \\
\Name{Yizirui Fang}$^{2}$ \Email{yfang52@jhu.edu} \\
\Name{Shan Zhang}$^{1}$ \Email{zhangshan@ufl.edu} \\
\Name{Seiyon M. Lee}$^{1}$ \Email{leeseiyon@ufl.edu} \\
\Name{Yiming Wang}$^{3}$ \Email{wang.yiming8@northeastern.edu} \\
\Name{Mark Trexler}$^{2}$ \Email{mtrexler@jhu.edu} \\
\Name{Anthony F. Botelho}$^{1}$ \Email{abotelho@coe.ufl.edu} \\
\addr $^1$ University of Florida, Gainesville, FL, USA \\
\addr $^2$ Johns Hopkins University, Baltimore, MD, USA \\
\addr $^3$ Northeastern University, Seattle, WA, USA
}

\begin{document}

\maketitle

\begin{abstract}
Integrating Large Language Models (LLMs) in educational technology reveals unprecedented opportunities to improve instructional design (ID), yet current approaches often prioritize automation over pedagogical rigor and human agency. This paper introduces ARCHED (AI for Responsible, Collaborative, Human-centered Education Instructional Design), a framework that implements a structured multi-stage workflow between educators and AI. Unlike existing tools that generate complete instructional materials autonomously, ARCHED cascades the development into distinct stages, from learning objective formulation to assessment design, each guided by Bloom's taxonomy and enhanced by LLMs. This framework employs multiple specialized AI components that work in concert: one generating diverse pedagogical options, another evaluating their alignment with learning objectives while maintaining human educators as primary decision-makers. ARCHED addresses critical gaps in current AI-assisted instructional design regarding transparency, pedagogical foundation, and meaningful human agency through this approach. This research advances the responsible integration of AI in education by providing a concrete, theoretically grounded framework that prioritizes human expertise and educational accountability.
\end{abstract}

\section{Introduction}
\label{sec:introduction}

The rise of Large Language Models (LLMs) has significantly changed educational technology, especially in instructional design \citep{choiUtilizingGenerativeAI2024}. As \citet{hodgesInnovationInstructionalDesign2024} observe, this transformation presents both unprecedented opportunities and significant challenges for educators. While LLMs offer powerful capabilities for content generation and assessment creation, their current implementation in educational settings often lacks pedagogical foundations and meaningful human oversight.

Recent studies have revealed concerning trends in AI-assisted instructional design. \citet{parsonsCanChatGPTPass2024} demonstrate that current AI tools can generate seemingly competent instructional materials, yet often fail to incorporate crucial pedagogical considerations. This automation-first approach risks diminishing the role of human expertise in educational design. Furthermore, \citet{sridharHarnessingLLMsCurricular2023} highlight that many existing systems operate as black boxes, making it difficult for educators to understand and validate the pedagogical reasoning behind generated content.

These challenges manifest in three critical areas. (1) The opacity of AI decision-making processes compromises educators' ability to ensure alignment with established educational frameworks. (2)  Current emphasis on full automation marginalizes human expertise, potentially undermining the quality and effectiveness of instructional design. (3) Standardization of AI-generated assessments often results in a narrow range of evaluation methods, limiting opportunities for innovative and diverse learning experiences \citep{chengTreeQuestionAssessingConceptual2024}.

This paper addresses these challenges through ARCHED, which fundamentally reimagines human-AI collaboration in instructional design by leveraging the capabilities of modern LLMs to enhance, rather than replace, human expertise. Specifically, we investigate the following research questions:

\textbf{RQ1:} How can AI-assisted instructional design systems maintain transparency while supporting efficient content development?

\textbf{RQ2:} What mechanisms enable meaningful integration of human expertise in AI-supported learning objective generation?

\textbf{RQ3:} How can AI tools promote diversity in assessment design while maintaining alignment with learning objectives?

Our work makes several contributions to the field. First, we introduce a transparent workflow that explicitly connects AI-generated content with established pedagogical frameworks through a structured multi-stage process. Second, we demonstrate how specialized AI components working in concert can enhance rather than diminish educator expertise, as validated through our empirical evaluation. Finally, we provide concrete evidence for how structured human-AI collaboration can diversify assessment strategies while maintaining pedagogical rigor, addressing a critical gap in current educational technology research.

\section{Related Work}
\label{sec:related_work}

The integration of Large Language Models (LLMs) in educational technology shows a transformative shift in instructional design practices. Our analysis of existing research reveals both significant advances and critical limitations in four interconnected domains.

\par{\textbf{Evolution of AI-Assisted Instructional Design.}} The development of AI-assisted instructional design evolves from basic automation to increasingly sophisticated pedagogical attempts. \citet{choiUtilizingGenerativeAI2024}'s SWOT analysis discloses a fundamental tension: while AI tools offer unprecedented efficiency in content creation, they require significant domain expertise for quality outcomes. Recent frameworks such as GAIDE \citep{dickeyGAIDEFrameworkUsing2024} and the design-build-test-learn approach \citep{chanAIAssistedEducationalDesign2024} demonstrate improved efficiency, but the challenges in maintaining pedagogical rigor persist. \citet{huTeachingPlanGeneration2024}'s evaluation of GPT-4 demonstrates that, while systems can produce structurally sound teaching plans, they struggle with context-specific pedagogical requirements. This limitation extends to graduate-level materials, where \citet{parsonsCanChatGPTPass2024} found that AI-generated content often lacks deep pedagogical understanding despite surface-level competence. \citet{dacostaEnhancingInstructionalDesign2024} and \citet{dacostaInvestigatingMediaSelection2024} further reveal that standardization in media selection and assessment methods may compromise educational effectiveness.

\par{\textbf{Learning Objectives and Assessment Generation.}}
\citet{sridharHarnessingLLMsCurricular2023} struggled with maintaining consistent taxonomic levels and authentic course alignment. Assessment generation faces similar challenges, with \citet{chengTreeQuestionAssessingConceptual2024}'s TreeQuestion system revealing how AI-generated assessments tend toward conventional formats, limiting evaluation diversity. This observation aligns with \citet{doughtyComparativeStudyAIGenerated2024}'s findings of significant quality variations between AI-generated and human-crafted programming assessments. \citet{rouzegarGenerativeAIEnhancing}'s multi-role AI assessment generation and \citet{linHowCanImprove2024}'s explanatory feedback highlighting demonstrate the performance of AI while revealing limitations in handling nuanced, domain-specific responses. 

\par{\textbf{Human-AI Collaboration Models in Education.}} 
\citet{almasreDevelopmentEvaluationCustom2024}'s GPT models show a consistent superiority of integrating human expertise over fully automated approaches, supporting \citet{linTeachersPerceptionsTeaching2022}'s necessitating pedagogical agency in utilizing AI capabilities. Innovative approaches include \citet{krushinskaiaDesignDevelopmentCoinstructional}'s co-instructional designer bot and \citet{liChatGPTGoodMiddle2024}'s ChatGPT in middle school instruction. While \citet{moussaReliableUtilizationInstructional2024} and \citet{madunicApplicationChatGPTInformation2024} structured approaches to GPT integration, \citet{lanTeachersAgencyEra2024} argue current frameworks often fail to fully address the complex interplay between human expertise and AI capabilities.

\par{\textbf{Transparency and Accountability Challenges.}}
\citet{parkAnalysisStateInstructional2024} systematically reveals widespread challenges in understanding and validating AI decision-making, while \citet{doamaralReflectionUseGenerative2024} highlights the need for transparent processes to maintain academic integrity while fostering innovation. \citet{hatmantoEmpoweringCreativeEducation2024} demonstrates how lack of transparency could disengage educator engagement and compromise pedagogy.
\citet{yadavScalingEvidencebasedInstructional2023}'s work on scaling evidence-based instructional design expertise through LLMs emphasizes the importance of maintaining clear connections to educational research and best practices while listing the ongoing challenge of accountability in AI-generated content.

\section{The ARCHED Framework: A Human-Centered Approach to AI-Assisted Instructional Design}
\label{sec:framework}

This section presents ARCHED (AI for Responsible, Collaborative, Human-centered Education Instructional Design), a framework that fundamentally reimagines the integration of AI in instructional design. Rather than pursuing full automation, ARCHED establishes a structured collaborative workflow that maintains human agency while leveraging AI capabilities to enhance the development of educational content. 

\subsection{Theoretical Foundation and Design Principles}

The development of ARCHED is grounded in the recognition that effective instructional design requires both pedagogical expertise and creative insight—qualities that neither AI systems nor humans alone can fully provide. Our framework builds upon Bloom's taxonomy as its pedagogical foundation, incorporating this established framework into a novel human-AI collaborative workflow. This approach addresses a fundamental limitation in current AI-assisted instructional design systems: their tendency to marginalize human expertise in favor of automation.

Three core principles guide our framework's design: First, human agency must be maintained throughout the instructional design process. Rather than generating complete instructional materials autonomously, AI systems should serve as collaborative tools that enhance human decision-making. Second, the framework must provide transparent reasoning for all AI-generated suggestions, enabling educators to make informed decisions about content adoption or modification. Third, the system should promote diversity in both learning objectives and assessment strategies, moving beyond the standardized approaches often favored by current AI systems.

\subsection{Framework Architecture and Implementation}
\begin{figure}[h]
\centering
\includegraphics[width=0.95\textwidth]{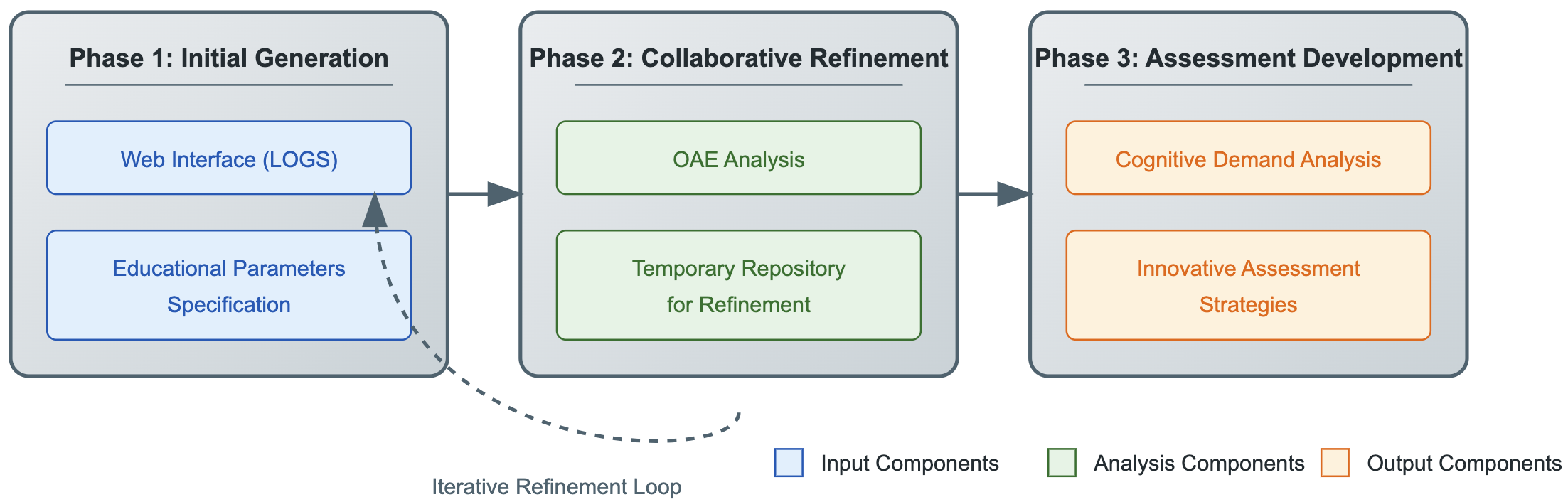}
\caption{ARCHED Implementation Methodology: This shows the three-phase workflow: (a) Initial generation through LOGS with educator input, specifying parameters such as grade level, subject area, and desired Bloom's taxonomy levels (b) Collaborative refinement with OAE analysis, and (c) Assessment development based on finalized objectives. Human educators maintain control throughout the process.}
\label{fig:framework}
\end{figure}

ARCHED implements a multi-stage collaborative workflow that integrates two specialized AI components: a Learning Objective Generation System (LOGS) and an Objective Analysis Engine (OAE). As shown in Figure \ref{fig:framework}, these components work in concert with human educators through three distinct phases while maintaining pedagogical rigor.

LOGS serves as the primary interface for the development of learning objectives. Unlike existing systems that attempt to generate complete sets of learning objectives autonomously, LOGS implements a staged approach that begins with educator input. Educators specify fundamental parameters including grade level, subject area, and desired Bloom's taxonomy level. This initial human input ensures that all generated content aligns with the specific educational context and learning goals.

OAE provides analytical support by evaluating generated objectives against established pedagogical criteria. This separation of generation and evaluation capabilities represents a key innovation in our framework. Using distinct AI systems for these tasks, we maintain clearer accountability and enable more robust analysis. OAE generates detailed reports that highlight potential improvements while maintaining explicit connections to pedagogical principles, enabling educators to make informed decisions about objective refinement.

\subsection{Technical Implementation}

We have developed an open-access web-based tool\footnote{A publicly available preview version of the tool can be accessed at \url{https://logen.viablelab.org/}.} that implements ARCHED through multiple specialized AI agents. The system employs careful prompt engineering and chain-of-thought reasoning, with LOGS using structured prompts incorporating Bloom's taxonomy for objective generation and OAE utilizing step-by-step reasoning chains for evaluation.

\begin{figure}[h]
\centering
\begin{minipage}[t]{0.48\textwidth}
\centering
\includegraphics[width=\linewidth]{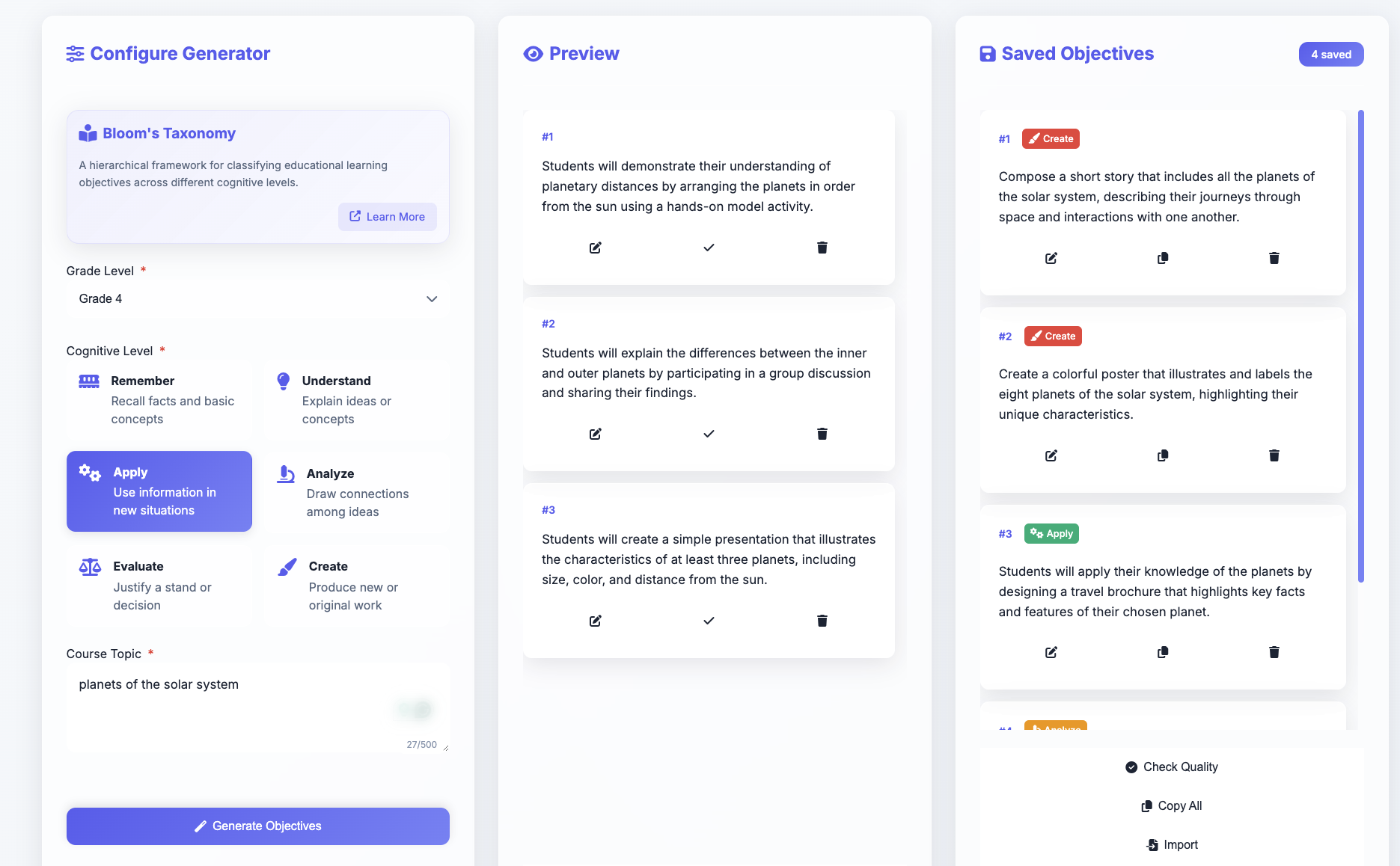}
\scriptsize
\caption{Main interface showing the three-phase workflow integration with continuous educator input in the loop}
\label{fig:interface1}
\end{minipage}
\hfill
\begin{minipage}[t]{0.48\textwidth}
\centering
\includegraphics[width=\linewidth]{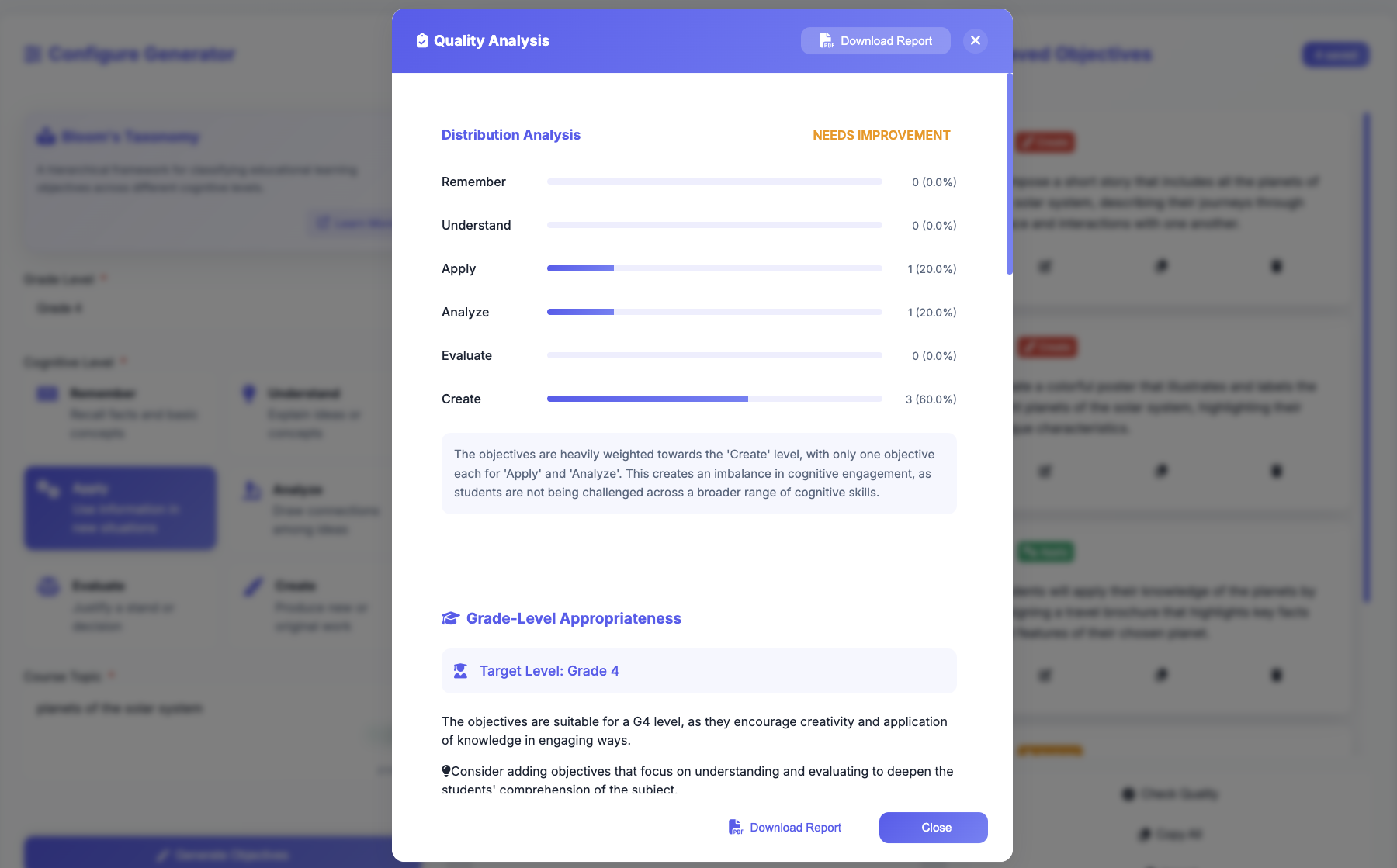}
\scriptsize
\caption{OAE analysis interface providing detailed pedagogical feedback and guidance}
\label{fig:interface2}
\end{minipage}
\end{figure}

Based on OpenAI's GPT-4O series APIs, the modular architecture of the system supports integration with various language models, including local deployments. Through an intuitive interface (Figure \ref{fig:interface1}), educators define course parameters and learning goals while receiving detailed feedback from OAE (Figure \ref{fig:interface2}). We initially tested both \texttt{GPT-4o} and \texttt{GPT-4o mini} and found their generation quality to be largely comparable. To enhance accessibility and ensure feasibility in resource-limited educational contexts, we primarily use \texttt{gpt-4o-mini-2024-07-18}, which offers a cost-effective alternative. Additionally, we explore open-weight LLMs as potential solutions, though barriers such as hardware acquisition, server hosting, and API costs remain challenges for widespread adoption. We continue to evaluate the most sustainable deployment strategies that balance affordability, accessibility, and pedagogical effectiveness.

\section{Results}
\label{sec:results}

We evaluated ARCHED through (1) a technical assessment of its learning objective analysis and generation capabilities and (2) preliminary expert feedback on its practical utility.

\subsection{Technical Evaluation}

To assess the framework's analytical capabilities, we tested it against 120 existing learning objectives from computer science education, with each objective independently classified by an educational expert according to Bloom's taxonomy. We employed weighted Cohen's Kappa (\(\kappa_w\)) for evaluation, using an ordinal weighting scheme to account for the hierarchical nature of taxonomic levels. The weighting scheme assigned 1.0 for perfect agreement, 0.8 for one-level difference, 0.6 for two-level difference, 0.4 for three-level difference, 0.2 for four-level difference, and 0.0 for five-level difference. For example, confusing adjacent levels (e.g., "Create" with "Evaluate") was penalized less than distant levels (e.g., "Create" with "Remember").

\begin{figure}[h]
\centering
\begin{minipage}{0.48\textwidth}
\centering
\includegraphics[width=\linewidth]{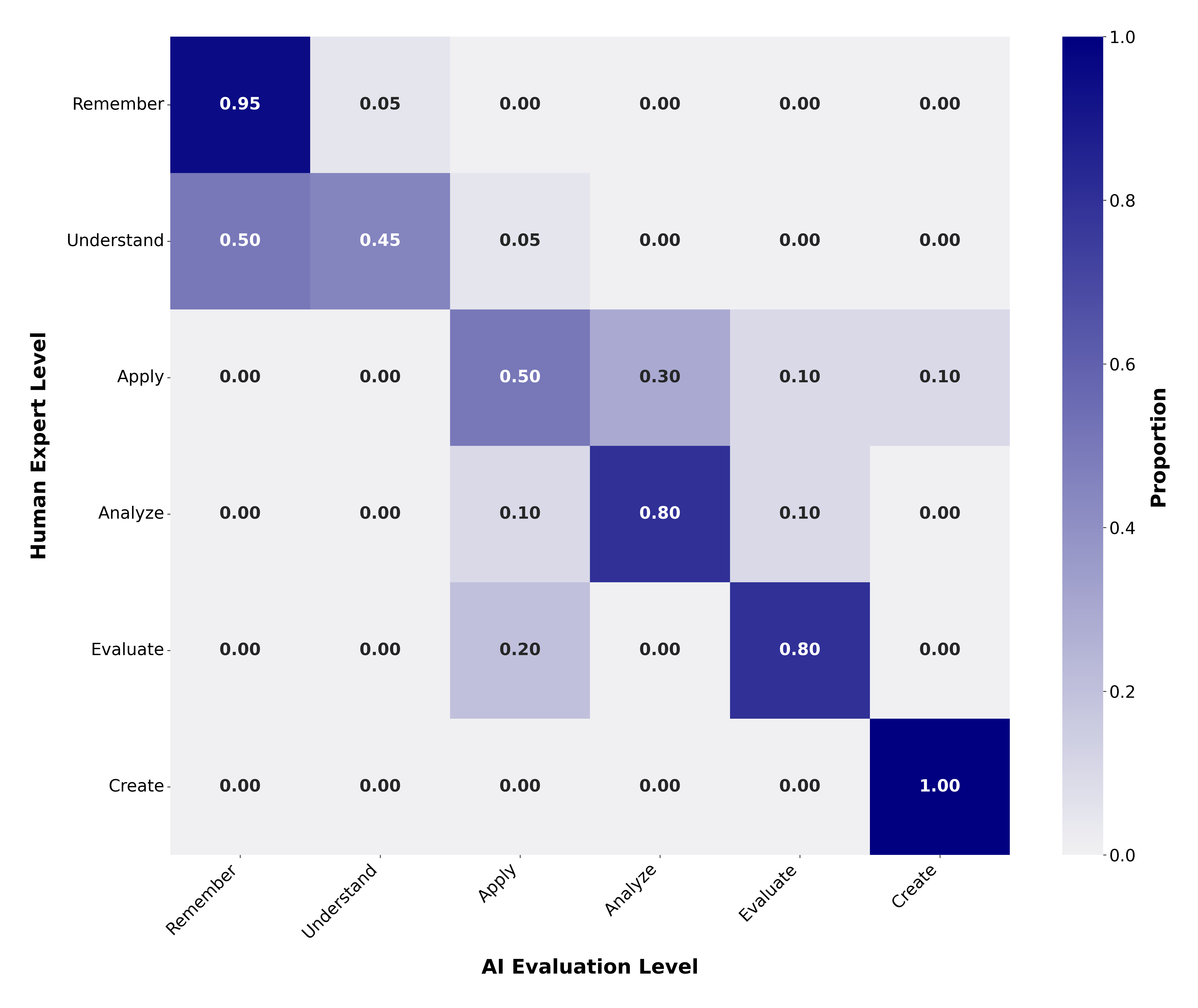}
\caption{Confusion matrix showing agreement between AI and expert classifications across Bloom's taxonomy levels}
\label{fig:confusion}
\end{minipage}
\hfill
\begin{minipage}{0.48\textwidth}
\centering
\scriptsize
\begin{tabular}{@{}lcc@{}}
\toprule
\small \textbf{Criterion} & \small \textbf{ARCHED} & \small \textbf{Human} \\
\midrule
\small Structural & \small 4.1±0.4 & \small 4.2±0.4 \\
\small Taxonomic & \small 4.0±0.5 & \small 4.1±0.4 \\
\small Measurable & \small 3.9±0.5 & \small 4.0±0.5 \\
\small Clarity & \small 4.0±0.4 & \small 4.1±0.4 \\
\small Technical & \small 3.8±0.6 & \small 4.0±0.5 \\
\bottomrule
\end{tabular}
\captionof{table}{Quality scores (mean\(\pm\)SD), with a maximum score of 5}
\label{tab:quality-scores}
\end{minipage}
\end{figure}

The framework achieved \(\kappa_w\) = 0.834 (95\% CI: [0.771, 0.891]), indicating strong agreement with expert classification. The confusion matrix analysis (Figure \ref{fig:confusion}) reveals several key patterns: strongest performance at the extremes of the taxonomy (Remember and Create levels), high accuracy in distinguishing higher-order thinking skills (Analysis and Evaluation), and expected confusion patterns between adjacent cognitive levels. This confusion primarily occurs in the middle range of the taxonomy, where cognitive processes naturally overlap, particularly between Understanding and its adjacent levels. These patterns align with known challenges in cognitive-level discrimination, even among human experts.

For generation quality assessment, we conducted a blind comparative analysis of 30 ARCHED-generated and 30 human-created learning objectives in computer science, mathematics, and physics. An expert evaluator assessed all objectives using five criteria on a scale of 1-5: structural completeness (ABCD and SMART model) \citep{heinich1996instructional,doran1981there}, taxonomic alignment, measurability, content clarity, and technical accuracy. Mann-Whitney U tests showed no significant differences between ARCHED and human-created objectives across all criteria (p \(>\) 0.05, Bonferroni-corrected), demonstrating the framework's ability to generate pedagogically sound objectives comparable to those created by experienced educators.

\subsection{Pilot User Feedback}

Two educational experts provided preliminary feedback on ARCHED's implementation, offering insights into how the framework's core principles manifest in practical instructional design workflows.

The experts' observations validated the framework's fundamental design principle of maintaining human agency through iterative refinement. The three-panel interface design, which implements ARCHED's staged workflow, was noted for effectively supporting this principle. As reflected in one observation, \textit{``The ability to adjust parameters and regenerate objectives based on specific needs is particularly useful. The interface makes it easy to fine-tune grade levels and cognitive levels until the objectives match course requirements.''} This feedback demonstrates how the framework's multi-stage workflow successfully enables meaningful human-AI collaboration.

The selective curation feature, a key implementation of ARCHED's human-centered design principle, was highlighted as essential for maintaining educator autonomy. \textit{``Having the freedom to select which objectives align with student needs and course goals, rather than accepting all generated content, helps maintain instructional control while leveraging AI assistance.''} noted an expert. This aligns with the framework's core goal of positioning AI as a supportive tool rather than an autonomous solution.

The quality analysis functionality, which implements ARCHED's emphasis on pedagogical rigor and transparency, received positive attention. The integration of Bloom's taxonomy into the analysis process, as specified in the framework design, was noted as particularly valuable. As an expert observed, \textit{``The visualization of the distribution analysis and the downloadable reports make it easy to identify gaps in cognitive coverage and document the development process.''} This feedback validates the framework approach to maintain the pedagogical alignment through explicit evaluation criteria.

The import feature for existing learning objectives demonstrated the flexibility of the framework in supporting various instructional design approaches. An observation highlighted that \textit{``The ability to upload and analyze existing objectives streamlines the review process, though integration with learning management systems would make it even more valuable."} This capability exemplifies ARCHED's design principle of enhancing rather than replacing existing educational practices.

This initial feedback suggests that ARCHED's implementation successfully realizes its core design principles of human agency, pedagogical rigor, and transparent AI assistance. The practical application of the framework demonstrates its potential to improve instructional design processes while maintaining educator control and pedagogical quality.

\section{Discussion}
\label{sec:discussion}
The development and implementation of ARCHED represents a significant advancement in AI-assisted instructional design by introducing a structured, human-centered framework that maintains pedagogical rigor while leveraging AI capabilities. Our evaluation results demonstrate how the framework's core design principles and components effectively address the key challenges identified in this work.

Regarding RQ1 on maintaining transparency in AI-assisted instructional design, ARCHED's multi-agent architecture fundamentally transforms the typical ``black box'' approach. By separating objective generation (LOGS) and evaluation (OAE) into distinct components, each with explicit reasoning processes, the framework provides educators with clear insight into the AI's decision-making process. The high rates of agreement (\(\kappa_w\) = 0.834) between OAE analysis and expert evaluations validate this approach, demonstrating that transparent AI systems can maintain high performance while supporting informed decision-making. This aligns with \citet{parkAnalysisStateInstructional2024}'s vision for explicable AI systems in education while addressing \citet{sridharHarnessingLLMsCurricular2023}'s concerns about opacity in AI-assisted instruction.

The collaborative workflow staged within the framework directly addresses RQ2 by reimagining how human expertise integrates with AI capabilities. Rather than attempting to automate the entire process, ARCHED implements specific interaction points where educator input guides AI generation and refinement. This design choice is validated by our comparative analysis, where ARCHED-generated objectives achieved quality scores comparable to human-created ones on all evaluation criteria. The framework's ability to maintain pedagogical sophistication while preserving educator agency demonstrates the effectiveness of our approach, supporting \citet{linTeachersPerceptionsTeaching2022}'s emphasis on meaningful human involvement in AI-assisted education.

For RQ3, concerning assessment diversity and objective alignment, our analysis focused primarily on the framework's ability to maintain pedagogical alignment in learning objective generation. The high-quality scores across structural completeness (4.1±0.4), taxonomic alignment (4.0±0.5), and measurability (3.9±0.5) demonstrate ARCHED's capability to generate well-aligned learning objectives that support diverse assessment possibilities. Although our current evaluation concentrated on the generation and alignment of objectives, the framework lays the foundation for the future development of assessment strategies that maintain this demonstrated pedagogical rigor. This addresses part of \citet{chengTreeQuestionAssessingConceptual2024}'s concerns about assessment alignment, though further work is needed to fully explore the framework's potential indirect assessment generation.

Although our evaluation demonstrates ARCHED's effectiveness, particularly in computer science education contexts, several limitations warrant consideration. Following \citet{huTeachingPlanGeneration2024}, the ability of the framework to generate higher-order thinking objectives requires further validation across diverse disciplines. Additionally, as noted by \citet{krushinskaiaDesignDevelopmentCoinstructional}, system adoption may face challenges in resource-constrained settings due to familiarization requirements.

Future research will focus on increasing the capacity of ARCHED in several domains and improving interaction with current learning management systems \citep{moussaReliableUtilizationInstructional2024}. These developments aim to create a more accessible and comprehensive tool for AI-assisted instructional design while maintaining our commitment to transparency, human agency, and pedagogical quality.

\acks{We would like to thank the National Science Foundation (2331379, 1903304, 1822830), Institute of Education Sciences (R305B230007), Schmidt Futures, Bill and Melinda Gates Foundation (080555, 078981), and OpenAI.}

% \bibliography{references}

\end{document}